\documentclass{aankg}
\usepackage{graphics}

\newcommand{\beqn}{\begin{eqnarray}}
\newcommand{\eeqn}{\end{eqnarray}}

\begin{document}

   \thesaurus{06     
              (19.53.1;  
               19.63.1)} 
   \title{Possible evidence of quark matter in neutron star X-ray binaries}


   \author{Norman K. Glendenning \and
   Fridolin Weber
          }


\institute{Nuclear Science Division  \&
   Institute for Nuclear and Particle Astrophysics\\
     Lawrence Berkeley  National Laboratory,
	  MS: 70-319 \\ Berkeley, California 94720
	  }

 \date{\today}

\authorrunning{Glendenning \& Weber}
   \titlerunning{Possible evidence of quark matter}
   \maketitle

   \begin{abstract}
  We study the spin evolution of
  X-ray neutron stars in binary systems, which are being spun up by mass
  transfer from accretion disks.  Our investigation reveals that a
  quark phase transition resulting from the changing central density
  induced by the changing spin, can lead to a pronounced peak
  in the  frequency distribution of X-ray neutron stars.
  This finding provides one of several  possible explanations available in the
  literature, or at least a contributor   to part of  the observed
   anomalous frequency distribution of neutron stars in low-mass X-ray
  binaries (LMXBs), which  lie in a  narrow band centered at about 300 Hz,
  as found by the Rossi Explorer (RXTE).

   \end{abstract}

\section{Role of quark phase transition}\label{sec:deco}

The density and pressure in the interior of neutron stars is high in comparison
with nuclear density by a factor of some 5 to 10 depending on the particular
models used to estimate it. At such densities
it is quite plausible that the quark constituents of hadrons loose their
association with particular hadrons---the deconfined quark matter phase
replaces the normal phase. In a nonrotating star, the radial boundaries between
quark core, mixed phase, and normal hadronic phase would remain fixed. However
in a rotating star, because of the centrifugal distortion 
of the density 
in the interior, these boundaries will change as the rotational frequency of
the star changes with time.

A structural change occurs in such a star with changes in
frequency (\cite{glen97:a}).  If there were no change in the nature of matter,
the stell\-ar fluid would res\-pond simply under the action of the cent\-rifugal
force. However, the compressibility of the normal nuclear matter phase and the
deconfined and relatively free Fermi gas of the quark matter phase, are
different.  The former must be less compressible than the latter. When a
ms pulsar spins down, its central density may rise above the critical
phase transition density and the
central core will then change
phase to softer quark matter; it is compressed both by its
own gravitational attraction, and by the weight of the overlying nuclear phase.
The reverse will be true
in  spinup due to accretion.
In either case,
the distribution of mass, radius and moment of inertia
are changed by a phase transition
  beyond those changes that would 
take place in an immutable fluid under
the action of a changing centrifugal force.  

These ideas were applied to the spin-down of a ms pulsar
(\cite{glen97:a,glen97:e,weber99:topr}).
It was found that as the quark matter core grew in radial
extent, the moment of inertia decreased anomalously, and could even introduce
an era of spin-up lasting for $\sim 10^7$ years (\cite{glen97:e}). The response
of the moment of inertia to changes in spin  is very like the
so-called ``backbending'' in nuclei predicted by Mottelson and Valatin
(\cite{mottelson82:a}) and discovered many years ago
(\cite{johnson72:a,stephens72:a}).

Accreting X-ray neutron stars provide a very interesting contrast to the
spin-down of isolated ms pulsars. The X-ray stars are being spun up by
the accretion of matter from a low-mass, less-dense  white
dwarf companion. They are
presumably the link between the canonical pulsars with mean period of $0.7$ sec
and the ms pulsars (\cite{klis98:b,chakrabarty98:a}).

If the critical deconfinement density falls within the range spanned by
canonical pulsars, quark matter will already exist in them but may be ``spun''
out of X-ray stars as their frequency increases during accretion.  We can
anticipate that in a certain frequency range, the changing radial extent of the
quark matter phase will actually inhibit changes in frequency because of the
increase in moment of inertia occasioned by the gradual disappearance of the
quark matter phase.  Accreters will tend to spend a greater length of time in
the critical frequencies than otherwise. There will be an anomalous number of
accreters that appear at or near the same frequency. This is what was found
recently with the Rossi X-ray Timing Explorer (RXTE) (\cite{klis00:a}).

Presumably, accreters commence their evolution near the death
line of active canonical
pulsars
with frequencies of $\nu \sim 1$ Hz and end as ms pulsars with
$\nu \sim 200 {\rm~to~} 600$ Hz. 
The spinup evolution of an accreting star is a more complicated problem than
that of the spindown of an isolated ms pulsar of constant baryon
number. It is complicated by the accretion of matter ($\dot M > 10^{-10}
M_\odot~{\rm yr}^{-1}$), a changing magnetic field strength (from $B \sim
10^{12} {\rm~to~} \sim 10^8$~G), and the interaction of the field with the
accretion disk.

\section{Spin-evolution of accreting X-ray neutron stars}\label{sec:spin}

The change in moment of inertia as a function of rotational frequency caused by
accretion of matter is similar to that described by \cite{glen97:a} 
for
spindown of a ms pulsar because of magnetic dipole radiation.
However, there are  additional phenomena as just mentioned.  Generally,
a canonical pulsar will have evolved from birth with moderate rotational
frequency to the deathline.  At that time, the usual drag of the dipole
radiation will be eclipsed by the accretion phenomena.  The spin-up torque of
the accreting matter causes a change in the star's angular momentum according
to the relation (\cite{elsner77:a,ghosh77:a,lipunov92:book})
\begin{eqnarray}
{{dJ} \over {d t}} = {\dot M} {\tilde l}(r_{\rm m}) - N(r_{\rm c}) \,
.
\label{eq:dJdt}
\end{eqnarray}
This can be rewritten as a time evolution equation for the
angular velocity  $\Omega$ of the accreting star
 $(\equiv
J/I)$,
\begin{equation}
  I(t) {{d\Omega(t)} \over {d t}} = {\dot M} {\tilde l}(t) - \Omega(t)
    {{dI(t)}\over{dt}} - \kappa \, \mu(t)^2 \, r_{\rm c}(t)^{-3} \, ,
    \label{eq:dOdt.1}\label{spinevolution}
    \end{equation}
    with the following definitions:
The accretion rate is denoted by
 $\dot{M}$  ($G=c=1$)  and
\begin{eqnarray}
{\tilde l}(r_{\rm m}) = \sqrt{M r_{\rm m}} 
\label{eq:l}
\end{eqnarray}
is the angular momentum added to the star per unit mass of accreted
matter. The quantity $N$ stands for the magnetic plus viscous torque
term ($\kappa\sim 0.1$),
\begin{equation}
N(r_{\rm c}) = \kappa \, \mu^2 \, r_{\rm c}^{-3} \, ,
\label{eq:N}
\end{equation}
with $\mu \equiv R^3 B$ the star's magnetic moment. 
The
quantities $r_{\rm m}$ and $r_{\rm c}$ denote the radius of the inner edge of
the accretion disk and the co-rotating radius, respectively, and are given by
$(\xi \sim 1)$
\begin{equation}
  r_{\rm m} = \xi \, r_{\rm A} \, ,
\label{eq:r_m}
\end{equation}
and 
\begin{eqnarray}
r_{\rm c} = \left( M \Omega^{-2} \right)^{1/3} \, .
\label{eq:r_c}
\end{eqnarray} Accretion will be inhibited by a centrifugal barrier if the
neutron star's magnetosphere rotates faster than the Kepler frequency at the
magnetosphere. Hence $r_{\rm m} < r_{\rm c}$, otherwise accretion onto the star
will cease.  The Alf\'en radius $r_{\rm A}$, where the magnetic energy density
equals the total kinetic energy of the accreting matter, in Eq.\ (\ref{eq:r_m})
is defined by
\begin{equation}
r_{\rm A} = \left( { {\mu^4} \over {2 M \dot{M}^2} } \right)^{1/7} \, .
\label{eq:r_A}
\end{equation}
We assume that the magnetic field evolves according to
\begin{equation}
B(t) = B(\infty) + (B(0) - B(\infty)) e^{-t/t_{\rm d}}
\end{equation} with $t=0$ at the start of accretion, and
where $B(0)=10^{12}$ G and $t_{\rm d} = 10^6$ yr.  
Such  a decay to an asymptotic value seems to be a feature of some treatments
of the magnetic field evolution (\cite{konar}).

It has previously been assumed that the moment of inertia 
in Eq.\ \ref{spinevolution} does not respond to
changes in the centrifugal force, and in that case, the above formula yields a
well-known estimate of the period to which a star can be spun up
(\cite{heuvel91:a}).  The approximation is true for slow rotation. However, the
response of the star to rotation becomes important as the star is spun up by
accretion.  Not only do changes in the distribution of matter occur but
internal changes in composition occur also because of changes induced in the
central density by centrifugal dilution (\cite{glen97:a}); both changes effect
the moment of inertia and hence the response of the star to accretion.  In this
Letter we wish to follow the spin-evolution of the star, and so, must take such
refinements into account.

\begin{figure}[htb]
\begin{center}
\resizebox{6cm}{!}{\includegraphics{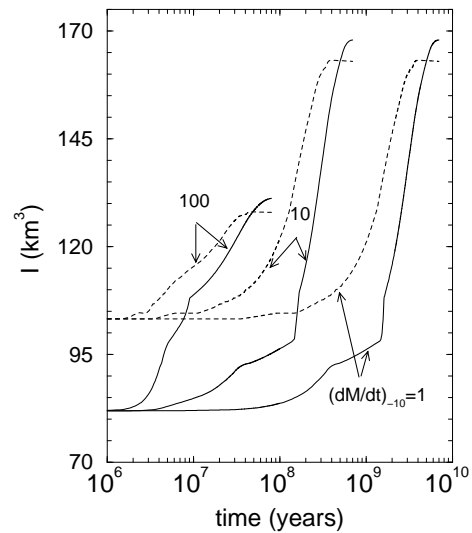}}
{ \caption { Moment of inertia of neutron stars with  (solid curves) and without
  (dashed curves) quark deconfinement  
assuming $0.4 M_\odot$ is accreted.)
\label{fig:It} 
}}
\end{center}
\end{figure}

The moment of inertia of ms
pulsars or of neutron star accreters has to be computed in GR
without making the usual assumption of
slow rotation (\cite{hartle67:a,hartle68:a}). Fortunately, we have previously
obtained an expression for the moment of inertia of a rotating star
(\cite{glen92:crust}).  
The expression is too cumbersome to reproduce
here.  
Stars that are
spun up to high  frequencies close to the breakup
limit (Kepler frequency)
undergo dramatic interior changes;  the
central density may change by a factor of four or so over that of a slowly
rotating star if a phase change occurs
during spin-up
(\cite{glen97:e,weber99:topr}).

Figure \ref{fig:It} shows how the moment of inertia changes for neutron stars
in binary systems that are spun up by mass accretion according to Eq.\ 
(\ref{eq:dOdt.1}) until  $0.4 M_{\odot}$
 has been accreted. The neutron star models are fully
described in (\cite{glen97:a}) and references therein,
and the initial mass of the star in our examples is $1.42 M_\odot$.
Confined nuclear matter is described by a covariant Lagrangian
describing the interaction of the members of the baryon octet with scalar,
vector and vector-isovector mesons and solved in the meanfield approximation. 
Quark matter is described by the MIT bag model.
In one case, it is assumed that a phase transition to quark matter occurs, and
in the other that it does not. This accounts for the different initial moments
of inertia, and also, as we see, the response to spinup.
Three accretion rates are assumed, which range from
$\dot{M}_{-10}=1$ to 100.  These rates are in accord with observations made on
low-mass X-ray binaries (LMXBs) observed with the Rossi X-ray Timing Explorer
(\cite{klis00:a}). The observed objects, which are divided into Z sources and
A(toll) sources, appear to accrete at rates of $\dot{M}_{-10} \sim 200$ and
$\dot{M}_{-10} \sim 2$, respectively.

\begin{figure}[htb]
\begin{center}
\resizebox{6cm}{!}{\includegraphics{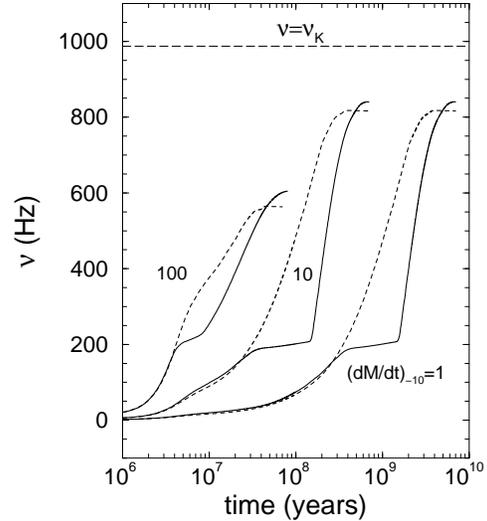}}
{ \caption {Evolution of spin frequencies ($\nu \equiv \Omega / 2\pi$)
of accreting neutron stars with (solid curves) and without (dashed
curves) quark deconfinement if $0.4 M_\odot$ is accreted. 
(If the mass of the donor star is less, then so is the maximum
attainable frequency.)
The spin saturation around $200$~Hz
signals the ongoing process of quark confinement in the stellar
centers. 
\label{fig:nue}
}}
\end{center}
\end{figure}

Figure \ref{fig:nue} shows the spin evolution of accreting neutron stars as
determined by the changing moment of inertia and the spin evolution equation
(\ref{spinevolution}).  Neutron stars without quark matter in their centers are
spun up along the dashed lines to equilibrium frequencies between about 600~Hz
and 850~Hz, depending on accretion rate and magnetic field. The $dI/dt$ term
for these sequences manifests itself only insofar as it limits the equilibrium
periods to values smaller than the Kepler frequency,
$\nu_{\rm K}$.  
In both Figs.\ \ref{fig:It} and  \ref{fig:nue} we assume that $0.4 M_{\odot}$
is accreted. Otherwise the maximum frequency attained is less.

The spin-up scenario is dramatically different for neutron stars in which quark
deconfinement occurs. In this case, as known from Fig.\ \ref{fig:It}, the
temporal conversion of quark matter into its mixed phase of quarks and confined
hadrons is accompanied by a pronounced increase of the stellar moment of
inertia. This increase contributes so significantly to the torque term
$N(r_{\rm c})$ in Eq.\ (\ref{eq:dOdt.1}) that the spin-up rate $d\Omega/dt$ is
driven to saturation around those frequencies at which the pure quark matter
core in the center of the neutron star gives way to the mixed phase of confined
hadronic matter and quark matter.  The star resumes ordinary spin-up if this
transition is completed.  The epoch during which the spin rates are saturated
are determined by attributes like the accretion rate, magnetic field, and its
assumed decay time. The epoch lasts between $\sim 10^7{\rm~and~}10^9$ yr
depending on the accretion rate at the values taken for the other factors.

\begin{figure}[htb]
\begin{center}
\resizebox{6cm}{!}{\includegraphics{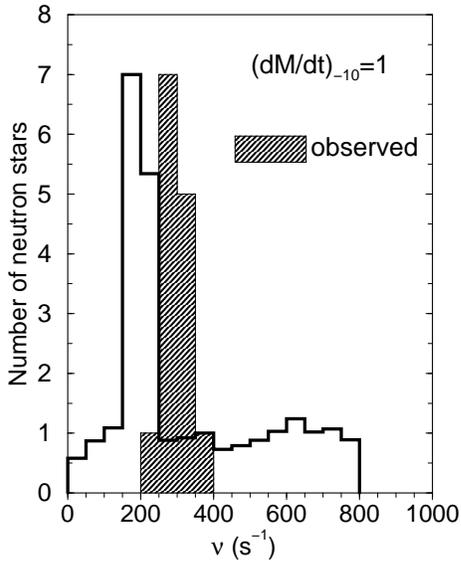}}
{ \caption { Frequency distribution of X-ray neutron stars. Calculated 
distribution is
normalized to the
number of observed objects (14). (The normalization causes a fractional number
to appear in many bins.)
 Data on  neutron stars in
LMXBs (A=Atoll sources, Z=Z sources) is from Ref.\  (\cite{klis00:a}).
The spike in the calculated distribution
corresponds to the spinout of the quark matter phase and the
corresponding growth of the moment of inertia.  Otherwise the spike
would be absent. \label{fig:bin}
}}
\end{center}
\end{figure}

We can translate the information in Fig.\ \ref{fig:nue} into a frequency
distribution of X-ray stars by
assuming that neutron stars begin their accretion
evolution at the average rate of one per million years. A different rate will
only shift some neutron stars from one bin to an adjacent one, but
will not change the basic form of the distribution.  The result is
shown in Fig.\ \ref{fig:bin}. The result is striking. Spinout of the
quark matter core as the neutron star spins up is signalled by a spike
in the frequency distribution. The concentration in frequency is
centered around 200 Hz, about 100 Hz lower than the observed spinup
anomaly. This discrepancy
is not surprising given the crude representation of confinement by the MIT bag 
model while the physics underlying the effect of a phase transition on
spin rate is robust.

We address now the bump in the histogram at high frequencies. 
Certainly there are high frequency {\sl pulsars}. However, if a histogram
of {\sl ms pulsar} frequencies is made from  
 Ref.\ (\cite{princeton}),
 a spike is found near 300 Hz, and a strong
attenuation in number of high 
frequency pulsars above the spike. So both the (sparse) data on X-ray
objects and on ms pulsars agree on a spike and on attenuation at high
frequency. Why the high frequency
bump (containing about 9 X-ray objects) in our Fig.\ \ref{fig:bin}? 
 The actual white dwarf masses
in these low-mass binaries is believed to be  $0.1$ to $0.4  M_\odot$.
We computed the frequency distribution for  donor  masses in this range
in steps of  $0.1  M_\odot$ and
until all mass has been transferred at the chosen rate.
(The result has little sensitivity to the accretion rate.)
Since we do not know the 
mass distribution of donors, we averaged the results. 
The highest frequencies are attributable  to the 
 $0.4  M_\odot$ mass donors. These presumably account for the
  high
  frequency
  tail of  the {\sl ms pulsar}  distribution.
  Of course it is only an assumption that the mass distribution of 
  donors is flat and that the donor is entirely consumed.
Each binary  represents
a unique combination of neutron star and companion masses, magnetic fields,
and accretion rates, but with unknown weight for these differences.  
We simply do not know how the high frequency end is attenuated, but it 
surely is, as observation tells us,  though not so severely as the
few data on LMXBs would suggest, inasmuch as they are believed to
be the pathway to ms pulsars, several
of which have frequencies  as high as
$\sim 650$  Hz.

\section{Summary}\label{ref:summary}

We have traced the time evolution of the moment of inertia and
rotational frequency for a neutron star accreting matter from a
low-mass companion, under various assumptions about the accretion rate
and for two stellar models, one an ordinary neutron star populated by
nucleons, hyperons and leptons, and one in which phase equilibrium
between ordinary and quark deconfined matter occurs within the density
range found in canonical pulsars.
In the second case
the computed frequency distribution of X-ray neutron stars
shows a spike, much as is observed in a recent compilation of data
(\cite{klis00:a}). There are various suggestions 
as to the
cause of the spike, several of which
we cite (c.f. \cite{bildsten,andersson,levin}). 
A possible contributing
mechanism which causes some accreters of suitable
mass  to resist spinup for a lengthy era is that
discussed in this paper---
the ongoing  reduction
of quark matter cores in the centers of neutron stars 
as they are spun up. 
This occurs because, with increasing spin, the density of the inner
region is centrifugally diluted until it falls 
below the threshold density at which quark matter
can exist, first in the center, and then in an expanding region.
As explained in the introduction, 
the conversion of quark mattter
to confined hadronic matter
manifests itself in an expansion of the
star and a significant increase in its
moment of inertia. As a consequence,
the angular momentum added to a neutron star during
this phase of evolution is then consumed by the star's expansion,
inhibiting a further spin-up until the quark matter has been converted
into a mixed phase of matter made up of hadrons and quarks.

\begin{acknowledgements}
This work was supported by the Deutsche Forschungsgemeinschaft (DFG),
and by the Director, Office of Science, Office of High Energy and
Nuclear Physics, Division of Nuclear Physics, of the U.S. Department
of Energy under Contract DE-AC03-76SF00098. 
 \end{acknowledgements}



\end{document}